\documentclass[12pt,letterpaper]{article}
\usepackage{amsmath,latexsym,amssymb,ifthen}
\usepackage{xspace}
\usepackage{verbatim}
\usepackage{enumerate}

\usepackage{graphicx,hyperref,color}

\def\red{\color{red}}
\def\tP_k{\tilde{P}_k}
\newcommand{\beq}[2]{\begin{equation}\label{#1}#2\end{equation}}

\newboolean{draft}
\setboolean{draft}{true}

\def\dD{{\vec D}}
\def\dE{\vec E}

\def\cT{\mathcal{T}}

\def\bu{{\bf u}}
\def\bv{{\bf v}}

\def\bz{{\bf z}}
\def\tC{{\tilde C}}
\newcommand{\set}[1]{\left\{#1\right\}}

\newcommand{\mults}[1]{\begin{multline*}#1\end{multline*}}

\def\cP{\mathcal{P}}

\newcommand{\proofstart}{{\noindent \bf Proof.\hspace{.2em}}}
\newcommand{\proofend}{\hspace*{\fill}\mbox{$\Box$}}

\def\cL{{\mathcal L}}

\def\cD{{\mathcal D}}

\newcommand{\lfrac}[2]{\left(\frac{#1}{#2}\right)}

\def\a{\alpha}
\def\b{\beta}

\def\e{\epsilon}
\def\f{\phi}

\def\g{\gamma}
\def\G{\Gamma}
\def\k{\kappa}

\def\z{\zeta}

\def\th{\theta}

\def\l{\lambda}

\def\p{\pi}

\def\r{\rho}

\def\S{\Sigma}

\def\om{\omega}

\newcommand{\mnote}[1]{\ifthenelse{\boolean{draft}}{\marginpar{\footnotesize\raggedright#1}}}{}

\newcommand{\rdup}[1]{\left\lceil #1 \right\rceil}

\newcommand{\whp}{w.h.p.}
\newcommand{\Whp}{W.h.p.\ }

\newcommand{\wh}[1]{\widehat{#1}}

\def\nn{\nonumber}

\usepackage{tikz}
\usetikzlibrary{arrows}
\usetikzlibrary{decorations}
\usetikzlibrary{shapes.misc}

\newcommand{\brac}[1]{\left( #1 \right)}

\newcommand{\expect}{\operatorname{\bf E}}
\def\E{\expect}
\renewcommand{\Pr}{\mathbb{P}} 
\newcommand\bfrac[2]{\left(\frac{#1}{#2}\right)}

\def\bC{\bar{C}}

\newtheorem{theorem}{Theorem}
\newtheorem{lemma}[theorem]{Lemma}
\newtheorem{corollary}[theorem]{Corollary}

\newtheorem{remthm}[theorem]{Remark}

\newcounter{thmtemp}
\newtheorem{proposition}[theorem]{Proposition}

\newenvironment{proof}{\proofstart}{\proofend}
\newcommand{\ignore}[1]{}
\newcommand{\nospace}[1]{}

\def\path{\operatorname{PATH}}

\def\ATSP{\ensuremath{\operatorname{ATSP}}\xspace}
\def\AP{\ensuremath{\operatorname{AP}}\xspace}

\def\m{\lambda}

\title{Solving a Random Asymmetric TSP Exactly in Quasi-Polynomial Time w.h.p.}

\author{
Tolson Bell\thanks{thbell@cmu.edu. Research supported in part by NSF Graduate Research Fellowship grant DGE 2140739.}~~and Alan M. Frieze\thanks{frieze@cmu.edu. Research supported in part by NSF grant DMS1952285}\\
\small Department of Mathematical Sciences\\[-0.8ex]
\small Carnegie Mellon University\\[-0.8ex]
\small Pittsburgh, PA, 15213, USA\\[-0.8ex]
}
\begin{document}
\maketitle\thispagestyle{empty}
\begin{abstract}\large
Let the costs $C(i,j)$ for an instance of the Asymmetric Traveling Salesperson Problem (ATSP) be independent copies of a non-negative random variable $C$ from a class of distributions that include the uniform $[0,1]$ distribution and the exponential mean 1 distribution with mean 1. We describe an algorithm that solves ATSP exactly in time $e^{\log^{2+o(1)}n}$, w.h.p. 
\end{abstract}
\section{Introduction}
Given an $n\times n$ matrix $(C(i,j))_{i,j\in[n]}$, the Asymmetric Traveling Salesperson Problem (ATSP) asks for the cyclic permutation $\pi$ on $n$ elements that minimizes $\sum_{i=1}^n C(i,\pi(i))$. We let $Z_{\ATSP}=Z_{\ATSP}^{(C)}$ denote the optimal cost for ATSP. As the ATSP is NP-hard in general, the goal of this paper is to analyze whether the ATSP can be solved efficiently when $(C(i,j))_{i,j\in[n]}$ is an average-case, rather than worst-case, matrix. We consider the situation where the costs $C(i,j)$ are independent copies of a continuous random variable $C$. We assume that $C$ has a density $f$ and satisfies
\begin{enumerate}[(i)]
\item $f(x)=a+bx+O(x^2)$ for $0\leq x\leq L$, where $a,b$ are constants and $aL\geq 1$. 
\item $\Pr(C\geq x)\leq \a e^{-\b x}$ for constants $\a,\b>0$.
\item To avoid some pathologies, we will also assume that there is a constant $M$ such that $f(x)\leq aM$ for $x\geq 0$.
\end{enumerate}
The prime examples are the uniform $[0,1]$ distribution $U[0,1]$ ($a=1,b=0,M=1$) and the exponential mean 1 distribution $EXP(1)$ ($a=1,b=-1,\a=\b=M=1$). The main result of this paper is as follows:
\begin{theorem}\label{th1}
Let the costs for ATSP, $(C(i,j))$, be independent copies of $C$, where $C$ has a distribution satisfying (i),(ii),(iii) above. There is an algorithm that solves ATSP exactly in $e^{\log^{2+o(1)}n}$ time, with high probability over the choices of random costs. 
\end{theorem}
\subsection{Background}
Given $(C(i,j))_{i,j\in[n]}$, we can define another discrete optimization problem. Let $S_n$ denote the set of permutations of $[n]=\set{1,2,\ldots,n}$ . The \emph{Assignment Problem} (\AP) is the problem of minimising $C(\p)=\sum_{i=1}^n C(i,\pi(i))$ over all permutations $\p\in S_n$, (while the \ATSP only optimizes over cyclic permutations). We let $Z_{\AP}=Z_{\AP}^{(C)}$ denote the optimal cost for the AP. 

Another view of the assignment problem is that it is the problem of finding a minimum cost perfect matching in the complete bipartite graph $K_{A,B}$ where $A=\set{a_1,a_2,\ldots,a_n}$, $B=\set{b_1,b_2,\ldots,b_n}$, and the cost of edge $(a_i,b_j)$ is $C(i,j)$.

It is evident that $Z_{\AP}^{(C)} \leq Z_{\ATSP}^{(C)}$. The \AP is solvable in  time $O(n^3)$ \cite{Tom71, EK72}. In 1971, Bellmore and Malone \cite{BeMa} conjectured that using the \AP in a branch and bound algorithm would give a polynomial expected time algorithm for the \ATSP. Lenstra and Rinnooy Kan \cite{LeR} and Zhang \cite{Zh} found errors in the argument of \cite{BeMa}.

Several authors, e.g.\ Balas and Toth~\cite{BalToth}, Kalczynski \cite{Kal}, Miller and Pekny \cite{MP}, Zhang \cite{Z} have investigated using the \AP in a branch-and-bound algorithm to solve the \ATSP and have observed that the \AP gives extremely good bounds on random instances. Experiments suggest that if the costs $C(i,j)$ are independently and uniformly generated as integers in the range $[0,L]$, then as $L$ gets larger, the problem gets harder to solve. Rigorous analysis supporting this thesis was given by Frieze, Karp and Reed \cite{FKR}. They showed that if $L(n)=o(n)$ then $Z_{\ATSP}=Z_{\AP}$ \whp\ and that \whp\ $Z_{\ATSP}>Z_{\AP}$ if $L(n)/n\to\infty${\red.}

We implicitly study a case where $L(n)/n\to\infty$. Historically, researchers have considered the case where the costs $C(i,j)$ are independent copies of the uniform $[0,1]$ random variable $U[0,1]$. This model was first considered by Karp \cite{K1}. He proved the surprising result that
\beq{Karp}{
Z_{\ATSP}-Z_{\AP}=o(1)\text{ w.h.p.}
}
Since w.h.p. $Z_{\AP}>1$, we see that this gives a rigorous explanation for the previous observation that the \AP often gives extremely good bounds on the \ATSP. Karp \cite{K1} proved \eqref{Karp} constructively, analysing an $O(n^3)$ {\em patching} heuristic that transformed an optimal AP solution into a good ATSP solution. Karp and Steele \cite{KS} simplified and sharpened this analysis, and Dyer and Frieze \cite{DF} improved the error bound through the analysis of a related more elaborate algorithm to $O\left(\frac{\log^4n}{n\log\log n}\right)$. Frieze and Sorkin \cite{FS} reduced the error bound to 
\beq{diff}{
Z_{\ATSP}-Z_{\AP}\leq\frac{\z\log^2n}{n}\text{ w.h.p.}
} 
 Frieze and Sorkin also used this result to give an $e^{n^{1/2+o(1)}}$ time algorithm to solve the \ATSP exactly w.h.p. The main result of this paper is that we significantly improve this run-time, reducing the exponent from $n^{1/2+o(1)}$ to $\log^{2+o(1)}n$. Our result also allows for a larger class of random distributions for $C$.

One might think that with such a small gap between $Z_{\AP}$ and $Z_{\ATSP}$, that branch and bound might run in polynomial time \whp\ Indeed one is encouraged by the recent results of Dey, Dubey and Molinaro \cite{DDM} and Borst, Dadush, Huiberts and Tiwari \cite{BD} that with a similar integrality gap, branch and bound with LP based bounds solves random multi-dimensional knapsack problems in polynomial time w.h.p. 
Given Theorem \ref{th1}, one is tempted to side with \cite{BeMa} and conjecture that branch and bound can be made to run in polynomial time w.h.p.

\subsection{Chernoff bounds}
We use the following Chernoff bounds on the tails of the binomial distribution $Bin(n,p)$: here $0\leq\e\leq1$.
\begin{align*}
\Pr(Bin(n,p)\leq (1-\e)np)&\leq e^{-\e^2np/2}.\\
\Pr(Bin(n,p)\geq (1+\e)np)&\leq e^{-\e^2np/3}.
\end{align*}
We also use McDiarmid's inequality: let $Z=Z(Y_1,Y_2,\ldots,Y_n)$ where $Y_1,Y_2,\ldots,Y_n$ are independent random variables. Suppose that 
$|Z(Y_1,Y_2,\ldots,Y_n)-Z(\wh Y_1,\wh Y_2,\ldots,\wh Y_n)|\leq c$ when $Y_i=\wh Y_i$ except for exactly one index. Then
\beq{mcdineq}{
\Pr(|Z-\E(Z)|\geq t)\leq \exp\set{-\frac{2t^2}{c^2n}}.
}  
\section{The Assignment Problem and Nearby Permutations}\label{sec2}
Frieze and Sorkin \cite{FS} proved that when the distribution of costs is $U[0,1]$, the following two lemmas hold w.h.p.:
\begin{lemma}\label{lem2a}
$\max_{e\in M^*}C(e)\leq\frac{\g\log n}{n}$ for some absolute constant $\g>0$.
\end{lemma}\begin{lemma}\label{lemdiff}
$Z_{\ATSP}-Z_{\AP}\leq\frac{\z\log^2n}{n}$ for some absolute constant $\z>0$.
\end{lemma}
In Section \ref{properties}, we will prove that these lemmas still hold for our more general class of distributions. For Lemma \ref{lemdiff}, this comes via a short coupling argument, but the analysis to generalize Lemma \ref{lem2a} is more involved. We will use $\g^*$ to denote $\frac{\g\log n}{n}$ and $\z^*$ to denote $\frac{\z\log^2n}{n}$ for the corresponding $\g$ and $\z$ under which we prove Lemmas \ref{lem2a} and \ref{lemdiff}.
\subsection{AP as a linear program}\label{trandom}
The assignment problem $\AP$ has a linear programming formulation $\cL\cP$. In the following $z_{i,j}$ indicates whether or not $(a_i,b_j)$ is an edge of the optimal solution.
\begin{equation}\label{LP}
\begin{split}
\cL\cP\qquad\text{Minimise } &\sum_{(i,j)\in [n]^2} C(i,j) z_{i,j}\\
\text{ subject to }&\sum_{j=1}^n z_{i,j}=1,\text{ for }i=1,2,\ldots,n.\\
&\sum_{i=1}^n z_{i,j}=1,\text{ for }j=1,2,\ldots,n.\\
&z_{i,j}\geq 0,\text{ for } (i,j)\in [n]^2.
\end{split}
\end{equation}
An optimal basis of $\cL\cP$ can be represented by a spanning tree $T^*$ of $K_{A,B}$ that contains the perfect matching $M^*$, see for example Ahuja, Magnanti and Orlin \cite{AMO}, Chapter 11. The $2n-1$ edges of $T^*$ are referred to as {\em basic} edges and the edges not in $T^*$ are referred to as non-basic edges. In the dual formulation, w.h.p., $T^*$ is consists of those $(i,j)\in[n]^2$ for which $u_i+v_j=C(i,j)$.

$\cL\cP$ has the dual linear program:
\begin{equation}\label{DLP} 
\begin{split}
\cD\cL\cP\qquad&\text{Maximise} \sum_{i=1}^n u_i + \sum_{j=1}^n v_j \\
&\text{ subject to }  u_i+v_j\leq C(i,j),\text{ for } (i,j)\in [n]^2.
\end{split}
\end{equation}
\begin{remthm}\label{remx}
Note that replacing $u_i,v_j$ by $u_i+\l,v_j-\l$ for all $i,j$ does not affect the constraints or the objective value. We can therefore, when necessary, choose an $s$ and  assume that $u_s=0$ for some $s\in[n]$. 
\end{remthm}
\begin{proposition} \label{unif-non-basic}
Condition on an optimal basis for \eqref{LP}, that is, fix $T^*$ (which is unique with probability 1) and fix $C(i,j)$ for all $(a_i,b_j)\in E(T^*)$, but do not yet fix $C(i,j)$ for $(a_i,b_j)\notin E(T^*)$. We may w.l.o.g.\ take $u_1=0$ in~\eqref{DLP}, whereupon with probability 1 the other dual variables are uniquely determined. Furthermore, the reduced costs of the {\em non-basic} variables $\bC(i,j) = C(i,j)-u_i-v_j$ are independently distributed as either (i) $C-u_i-v_j$ if $u_i+v_j<0$ or (ii) $C-u_i-v_j$ conditional on $C\geq u_i+v_j$, if $u_i+v_j\geq 0$. 
\end{proposition}
\begin{proof}
The $2n-1$ dual variables are unique with probability 1 because they satisfy $2n-1$ full rank linear equations. The only conditions on the non-basic edge costs are that $C(i,j)\geq (u_i+v_j)^+$, where $x^+=\max\set{x,0}$. 
\end{proof}
\section{Outline Proof of Theorem \ref{th1}}
Recall the complete bipartite graph $K_{A,B}$ where $A=\set{a_1,a_2,\ldots,a_n}$, $B=\set{b_1,b_2,\ldots,b_n}$, and the cost of edge $(a_i,b_j)$ is $C(i,j)$. Let $M^*$ denote the minimum-cost perfect matching in $K_{A,B}$, which is the solution to the \AP. Any other perfect matching of $K_{A,B}$ can be obtained from $M^*$ by choosing a set of vertex disjoint {\em alternating cycles} $C_1,C_2,\ldots,C_m$ in $K_{A,B}$ and replacing $M^*$ by $M^*\oplus C_1\cdots\oplus C_m$. Here an alternating cycle is one whose edges alternate between being in $M^*$ and not in $M^*$. We use the notation $S\oplus T=(S\setminus T)\cup (T\setminus S)$.

For a matching $M$ we let $C(M)=\sum_{e\in M}C(e)$. The basic idea of the proof is to show that if a matching $M$ is ``too different'' from $M^*$, then w.h.p. $C(M)-C(M^*)>\tfrac{\z\log^2n}{n}$ (where $\z$ is from \eqref{diff}), and thus by (\ref{diff}), $M$ cannot be the optimal \ATSP solution. Once we have shown this, it does not take too long to check all possible $M$ that are ``similar to'' $M^*$, to see if $M$ defines a tour and then determine its total cost.

More specifically, let $T^*\supseteq M^*$ be the spanning tree representing the optimal basis in the LP formulation of AP (see Section \ref{trandom}). Our definition of ``too different'' is that $M$ contains more than $\log^{2+o(1)}n$ edges outside $T^*$. Each such edge corresponds to a non-basic variable. These reduced costs turn out to be conditionally independent. This makes it easy to show that the sum of ``many'' of them is greater than $\tfrac{\z\log^2n}{n}$ w.h.p. Lemma \ref{FewEdgesChanged} shows that this will w.h.p. preclude $M$ from corresponding to the optimum to ATSP. This is because the sum of the reduced costs associated with these non-$T^*$ edges will be greater than the upper bound in \eqref{diff} w.h.p.

In Sections \ref{TB} and \ref{iteration}, we will prove Theorem \ref{th1} in the special case where the distribution of the costs $C(i,j)$ is exponential mean one, $EXP(1)$, i.e. $\Pr(C\geq x)=e^{-x}$. We need to make this assumption for the proof of Lemma \ref{steady}. In Section \ref{generalizedist}, we will subsequently generalise our proof to the full class of distributions in Theorem \ref{th1}.
﻿\section{Trees and bases}\label{TB}
We have that for every optimal basis $T^*$,
\beq{uivj}{
C(i,j)=u_i+v_j\text{ for }(a_i,b_j)\in E(T^*)
}
and
\beq{uivj1}{
C(i,j)> u_i+v_j\text{ for }(a_i,b_j)\notin E(T^*).
}
Define the $k$-neighborhood of a vertex to be the $k$ vertices nearest it, where distance is given by the matrix~$C$. Let the $k$-neighborhood of a set be the union of the $k$-neighborhoods of its vertices. In particular, for a complete bipartite graph $K_{A,B}$ and any $S \subseteq A,T\subseteq B$,
\begin{align} 
N_k(S) & =\{b\in B:\;\exists s\in S\text{ s.t.\ $(s,b)$ is one of the $k$ least cost edges incident with $s$} \},\label{NS}\\
N_k(T) & =\{a\in A:\;\exists t\in T\text{ s.t.\ $(a,t)$ is one of the
$k$   least cost edges incident with $t$} \}.\label{NT}
\end{align}
Given the complete bipartite graph $K_{A,B}$ and a perfect matching $M$, we define a directed graph $\dD_M$ as follows: it has \emph{backwards} matching edges $\dE_{M}$ and forward ``short'' edges $\dE_{\bar M}$, where
\begin{align*}
\dE_M &=\{(b,a): \; b \in B, \: a \in A, \: \set{a,b}\in M\}	.\\ 
\dE_{\bar M}&=\{(a,b): \; a\in A, \: b\in N_{40}(a)\}
 \cup \; \{(a,b): \; b\in B, \: a\in N_{40}(b)\} .
\end{align*}
Paths in $\dD_M$ are necessarily {\em alternating}, i.e. the edges  alternate between being in $M$ and not in $M$. We assign a cost $C(i,j)$ to edge $(a_i,b_j)\in \dE_{\bar M}$ and a cost $-C(i,j)$ to an edge $(b_j,a_i)\in \dE_M$. The cost of a path $P$ being the sum of the costs of the edges in the path equals the change in the cost of a matching due to replacing $P\cap \dE_M$ by $P\cap\dE_{\bar M}$. Lemma \ref{lem2ab} below proves that w.h.p. for every perfect matching $M$, $a\in A,b\in B$, 
\beq{EP}{
\text{$\dD_M$ contains a path $P$ from $a$ to $b$ of cost at most $\g^*$.}
}
Given this we can prove a high probability upper bound on the $u_i,v_j$.
\begin{lemma}\label{uvboundl}
Let \bu,\bv\ be optimal dual variables and suppose that $u_1=0$. Then,
\beq{uvbound}{
\text{$|u_i|,|v_i|\leq 2\g^*$ for $i\in [n]$, w.h.p.}
}
\end{lemma}
\begin{proof}
Fix $a_i,b_j$ and let $P=(a_{i_1},b_{j_1},\ldots,a_{i_k},b_{j_k})$ be the alternating path from $a_i$ to $b_j$ promised by Lemma \ref{lem2ab}. Then, using \eqref{uivj} and \eqref{uivj1}, we have
\beq{pathP}{
\g^*\geq C(P)=\sum_{l=1}^kC(i_l,j_l)-\sum_{l=1}^{k-1}C(i_{l+1},j_l)\geq \sum_{l=1}^k(u_{i_l}+v_{j_l})-\sum_{l=1}^{k-1}(u_{i_{l+1}}+v_{i_l})=u_i+v_j. 
}
For each $i\in [n]$ there is some $j\in [n]$ such that $u_i+v_j=C(i,j)$. This is because of  the fact that $a_i$ meets at least one edge of $T^*$ and we assume that \eqref{uivj} holds. We also know that from \eqref{pathP} that $u_{i'}+v_j\leq\g^*$ for all $i'\neq i$. It follows that $u_i-u_{i'}> C(i,j)-\g^*\geq -\g^*$ for all $i'\neq i$. Since $i$ is arbitrary, we deduce that $|u_i-u_{i'}|\leq\g^*$ for all $i,i'\in[n]$. Since $u_1=0$, this implies that $|u_i|\leq \g^*$ for $i\in [n]$. We deduce by a similar argument that $|v_j-v_{j'}|\leq \g^*$ for all $j,j'\in [n]$. Suppose that the optimal matching 
\[
M^*=\set{\set{a_i,b_{\f(i)}}:i\in [n]}.
\]
 Then we have $u_i+v_{\f(i)}=C(i,\f(i))$, and because (as will be shown in Section \ref{properties}) $C(i,\f(i))\leq \g^*$, we have $|v_j|\leq 2\g^*$ for $j\in[n]$.
\end{proof}

Condition on the edges of the matching $M^*$ and let $G_+$ denote the subgraph of $K_{A,B}$ induced by the edges $(a_i,b_j)$ for which $u_i+v_j\geq0$ where $\bu,\bv$ are optimal dual variables. Let $\cT_+$ denote the set of spanning trees of $G_+$ that contain the edges of $M^*$. Note that \bu,\bv\ do not determine $T^*$. Let $f(\bu,\bv)$ denote the joint density of $\bu,\bv$ then
\begin{lemma}\label{steady}
If $T\in \cT_+$ then
\beq{prop2}{
\Pr(T^*=T\mid\bu,\bv)f(\bu,\bv)= \prod_{(a_i,b_j)\in G_+}e^{-u_i-v_j},
}
which is independent of $T$.
\end{lemma}
\begin{proof}
Fixing \bu,\bv\ and $T$ fixes the lengths of the edges in $T$. If $(a_i,b_j)\notin E(T)$ then $\Pr(C(i,j)\geq u_i+v_j)=1$ if $u_i+v_j<0$ and $e^{-(u_i+v_j)} $ otherwise. Remember that we are assuming that $C$ is exponential mean one at the moment. If $(a_i,b_j)\in E(T)$ then the density of $C(i,j)$ is $e^{-(u_i+v_j)}$. Thus, Proposition \ref{unif-non-basic} and the ``memoryless property'' of the exponential distribution imply
\begin{align}
\Pr(T^*=T\mid\bu,\bv)f(\bu,\bv)&=\prod_{(a_i,b_j)\in G_+\setminus E(T)}e^{-(u_i+v_j)} \prod_{(a_i,b_j)\in E(T)}e^{-(u_i+v_j)}\nn\\
&=  \prod_{(a_i,b_j)\in G_+}e^{-(u_i+v_j)}.\label{produv}
\end{align}
In the first product we use \eqref{uivj1}, and the second product comes from \eqref{uivj} and from the density function of the costs.  
\end{proof}

Thus 
\beq{Tris}{
\text{$T^*$ is a uniform random member of $\cT_+$.}
}
Now let ${\wh G}_+$ be the multi-graph obtained from $G_+$ by contracting the edges of $M^*$ and let $\wh T^*$ be the corresponding contraction of $T^*$. 
\begin{lemma}\label{complete}
The distribution of the tree $\wh T^*$ is isomorphic to that of a random spanning tree of the graph $H=K_n\cup \wh M$ where w.h.p. ${\wh M}$ is a matching of size at most $\l^*=\log^5n$. (${\wh M}$ yields double edges, other edges occur once.)
\end{lemma}
\begin{proof}
We have that for all $i,j\in[n]$,
\beq{saved1}{
(u_i+v_{\f(j)})+(u_j+v_{\f(i)})=(u_i+v_{\f(i)})+(u_j+v_{\f(j)})=C(i,\f(i))+C(j,\f(j))>0.
}
So, either $u_i+v_{\f(j)}>0$ or $u_j+v_{\f(i)}>0$ which implies that ${\wh G}_+$ contains the edge $\set{a_i,a_j}$ and so ${\wh G}_+$ contains $K_A$, the complete graph on $A$, as a subgraph. So from \eqref{Tris}, $\wh T^*$ consists of a random spanning tree of $K_A$ plus a set of edges ${\wh M}$. The edges ${\wh M}$ arise when both $u_i+v_{\f(j)}>0$ and $u_j+v_{\f(i)}>0$ and where both edges have cost at most $4\g^*$. 

We know from \eqref{uivj} and Lemma \ref{uvboundl} that $\wh T^*$ only contains edges of cost at most $4\g^*$. Thus each repeated edge arises from a cycle of length 4 $(a_i,b_{\f(j)},a_j,b_{\f(i)},a_i)$  in which each edge has length at most $4\g^*$. The expected number of such cycles is $O((n\g^*)^4)$ and so by the Markov inequality, $|{\wh M}|\leq \log^5n$ w.h.p. At this density, any copies of $C_4$ will be vertex disjoint w.h.p., as can easily be verified by a first moment calculation. 
\end{proof}

\subsection{Alternating paths}
We call a path in $T^*$ a \emph{basic alternating path} if it is a single edge of $M^*$, or if it is an $M^*$-alternating path whose first and last edges belong to $M^*$.  After contracting the edges of $M^*$, every non-trivial basic alternating path admits a unique orientation whose image is a directed path in the digraph $D$ described below.

\begin{lemma}\label{alt}
Assuming the result of Lemma \ref{complete}, let $X_0=n$ and for $k\geq 1$ let $X_k$ be the number of basic alternating paths whose image in $\wh T^*$ has $k$ edges. Conditional on $\bu,\bv$ and $M^*$,
\[
\E(X_k\mid\bu,\bv,M^*)\leq n(k+1)\rho_n^k,
\]
where
\[
\rho_n=\frac{\brac{\binom n2+|\wh M|}^{1/2}}{n}=\frac1{2^{1/2}}+o(1).
\]
In particular, for all sufficiently large $n$ and $1\leq k\leq n-1$,
\[
\E(X_k\mid\bu,\bv,M^*)\leq n(k+1)\brac{\frac34}^k
\leq n^2\brac{\frac34}^k.
\]
\end{lemma}
\begin{proof}
Let $\cT$ be the set of spanning trees of $K_n$ and let $T_0$ be a uniformly chosen from $\cT$. For $e\in E(K_n)$, write $I_e=\mathbf 1_{\{e\in T_0\}}$.  Define a digraph $D$ on $[n]$ by orienting an edge $\set{i,j}$ of $H$ from $i$ to $ j$ whenever $(a_i,b_{\f(j)})\in G_+$.  Fix a directed simple path $\vec P$ of length $k$ in $D
$, and let $P$ be its underlying undirected path. Each arc of $\vec P$ is associated with one of the edge-copies of $H$. Let $\vec {\wh T^*}$ denote the orientation of $\wh T^*$ induced by the orientation of $D$. Then,
\begin{align*}
\Pr(\vec P\subseteq \vec{\wh T^*}\mid\bu,\bv,M^*)&=\frac{\sum_{T_0\in \cT}\mathbf 1_{\{P\subseteq T_0\}}\prod_{e\in \wh M\setminus P}(1+I_e)}{\sum_{T_0\in \cT}\prod_{e\in \wh M}(1+I_e)}\\
&=\frac{\E\brac{\mathbf 1_{\{P\subseteq T_0\}}\prod_{e\in\wh M\setminus P}(1+I_e)}}{
\E\brac{\prod_{e\in\wh M}(1+I_e)}}.
\end{align*}
The edge indicators of a uniform spanning tree are negatively associated; see \cite{LP}. Hence the numerator is at most
$\Pr(P\subseteq T_0)\,\E\brac{\prod_{e\in\wh M\setminus P}(1+I_e)}$. The denominator is at least the second factor in this display. Therefore
\[
\Pr(\vec P\subseteq \vec{\wh T^*}\mid\bu,\bv,M^*)\leq \Pr(P\subseteq T_0)=\frac{k+1}{n^k},
\]
where the last equality follows from the fact that there are exactly $(k+1)n^{n-k-2}$ spanning trees of $K_n$ contain the fixed path $P$.

Let $A$ be the adjacency matrix of $D$. The number of directed simple paths of length $k$ in $D$ is at most the number of directed walks of length $k$. Thus, writing $\mathbf e$ for the all-one vector and $||\cdot||_F$ for the Frobenius norm,
\[
N_k(D)\leq \mathbf e^{\mathsf T}A^k\mathbf e \leq n\|A\|_2^k \leq n\|A\|_F^k=n\brac{\binom n2+|\wh M|}^{k/2}.
\]
Multiplying this by $(k+1)/n^k$ gives
\[
\E(X_k\mid\bu,\bv,M^*)\leq n(k+1)\left(\frac{\brac{\binom n2+|\wh M|}^{1/2}}{n}\right)^k.
\]
Since $|\wh M|\leq\log^5n$, the term in parentheses is $1/\sqrt2+o(1)$ and is at most $3/4$ for all sufficiently large $n$.
\end{proof}

\begin{corollary}\label{cor1}
W.h.p. every basic alternating path has $O(\log n)$ edges.
\end{corollary}
\begin{proof}
Take $K=5\log n$. Conditional on the events in Lemma \ref{complete}, Lemma \ref{alt}, the Markov inequality gives
\[
\Pr(X_K>0\mid\bu,\bv,M^*)
\leq n(K+1)\brac{\frac34}^{K}=o(1).
\]
Every path whose contracted image has more than $K$ edges contains one with exactly $K$ edges. The uncontracted alternating path has at most twice as many edges, up to an additive constant.
\end{proof}

Fix $\om=\om(n)$ to be an arbitrary function such that $\om\to\infty,\om=\log^{o(1)}n$.

\begin{lemma}\label{basicpaths}
W.h.p. there are at most $m=\om n$ basic alternating paths, each using $O(\log n)$ edges.
\end{lemma}
\begin{proof}
Let $Z$ denote the total number of basic alternating paths. Conditional on the events in Lemma \ref{complete}, Lemma \ref{alt} we have, uniformly in $\bu,\bv$ and $M^*$,
\[
\E(Z\mid\bu,\bv,M^*)
\leq n+\sum_{k\geq1}n(k+1)\brac{\frac34}^k
\leq 16n.
\]
(The initial $n$ allows for paths that are contractions of an edge in $M^*$.) Consequently,
\[
\Pr(Z>\om n\mid\bu,\bv,M^*)\leq \frac{16}{\om}=o(1)
\]
by Markov's inequality. Combining this with Corollary \ref{cor1} proves the lemma.
\end{proof}

So, w.h.p. the matching $M$ corresponding to the ATSP solution is derived from a collection of short basic alternating paths $P_1,P_2,\ldots,P_m$ joined by non-basic edges to create alternating cycles $Q_1,Q_2,\ldots,Q_\ell$. Now consider an alternating cycle $Q=(a_{i_1},b_{j_1},\ldots,b_{j_t},a_{i_1})$ made up from such paths by adding non-basic edges joining up the endpoints. Putting ${\tilde C}(i,j)=C(i,j)-u_i-v_j$, we have that where $j_t=\f(i_t)$ for $t=1,2,\ldots,k$,
\begin{align*}
C(Q\oplus M^*)-C(M^*)&=\sum_{k=1}^t(C(i_{k+1},j_k)-C(i_k,j_k))\\ 
&=\sum_{k=1}^t(({\tilde C}(i_{k+1},j_k))+u_{i_{k+1}}+v_{j_k})-({\tilde C}(i_k,j_k)+u_{i_k}+v_{j_k})\\
&=\sum_{k=1}^t{\tilde C}(i_{k+1},j_k)\\
&=\sum_{\substack{k=1\\(i_{k+1},j_k)\text{ non-basic}}}^t{\tilde C}(i_{k+1},j_k).
\end{align*}
\begin{lemma}\label{FewEdgesChanged}
The optimal solution to the ATSP uses at most $\log^{2+o(1)}n$ non-basic edges w.h.p. 
\end{lemma}
\begin{proof}
For any $k\in\mathbb{N}$, let $Z_k$ be the number of perfect matchings $M$ in $K_{A,B}$ with $|M\setminus T^*|=k$ and $C(M)-C(M^*)\le\z^*$ (see Lemma \ref{lemdiff}).

For a perfect matching $M$ with $|M\setminus T^*|=k$, we have that $M\oplus M^*$ consists of $\ell$ cycles $Q_1,Q_2,\dotsc,Q_\ell$ where each $Q_i$ has $k_i$ edges outside $T^*$ where $k_1+\dotsc+k_\ell=k$. For each $i\in[\ell]$, we specify $Q_i$ by choosing $k_i$ alternating paths in $T^*$, then ordering and orienting them in at most $k_i!2^{k_i}$ ways. $M$ can correspond to the ATSP solution only if the $\tC$ of the $k$ corresponding non-basic edges sum to at most $\zeta^*$. Thus, if $\tC_1,\tC_2,\ldots,\tC_k$ are independent random variables distributed as (i), (ii) of Proposition \ref{unif-non-basic}, we have
\begin{align}
\E(Z_k)&\leq\sum_{\ell=1}^k\sum_{k_1+\cdots+k_\ell=k}\prod_{i=1}^\ell \binom{m}{k_i}k_i!2^{k_i}\Pr\brac{\tC_1+\tC_2+\cdots+\tC_k\leq \zeta^*}\label{1}\\
&\leq (2m)^k\frac{(2\zeta^*)^k}{k!}\sum_{\ell=1}^k|\set{k_1,\ldots,k_\ell\geq 2:k_1+\cdots+k_\ell=k}|\label{2}\\
&\lesssim (2m)^k\frac{(2\zeta^*)^k}{k!} \sum_{\ell=1}^k\binom{k}{\ell}\nn\\
&\leq \bfrac{8\z e\om\log^{2}n}{k}^k.\label{3}
\end{align}
Where, to go from \eqref{1} to \eqref{2}, we used
\begin{align*}
\Pr\brac{\tC_1+\cdots+\tC_k\leq \z^*}&\leq 
\int_{z_1+\cdots+z_k\leq \z^*}\prod_{i=1}^ke^{-z_i+4\g^*}d\bz\\
&\lesssim 2^k \int_{z_1+\cdots+z_k\leq\z^*}1d\bz=\frac{(2\z^*)^k}{k!}.
\end{align*}
The proof is completed by noting that $\E(\sum_{\log^{2+o(1)}n}^\infty Z_k)=o(1)$. (The $4\g^*$ in the first line comes from the $u_i+v_j$ in (i) of Proposition \ref{unif-non-basic}.)
\end{proof}

\section{Finishing the proof of Theorem \ref{th1}}\label{iteration}
Now that we know the solution to the ATSP satisfies Lemma \ref{FewEdgesChanged}, we will give an algorithm to iterate over the possible ATSP solutions. As w.h.p.~every potential ATSP solution uses at most $\log^{2+o(1)}n$ non-basic edges, and there are $n^2$ edges in total, we can quickly see that there are at most \[
\binom{n^2}{\log^{2+o(1)}n}\le(n^2)^{\log^{2+o(1)}n}\le e^{\log^{3+o(1)}n}
\]
potential ATSP solutions to iterate through. The goal of the technical work in this section is to describe an algorithm that works with a run-time reduced from $e^{\log^{3+o(1)}n}$ to $e^{\log^{2+o(1)}n}$ The intuitive reason that we can obtain run-time $e^{\log^{2+o(1)}n}$ is the following lemma, recalling $\z^*$ defined at the start of Section \ref{sec2}:
\begin{lemma}\label{AllTSPsmall}
W.h.p., every non-basic edge in the optimal solution to the ATSP that is not in $T^*$ has reduced cost at most $\z^*$.
\end{lemma}
\begin{proof}
Let $\tilde C_1,\tilde C_2,\ldots,\tilde C_k$ be the reduced costs of the non-basic edges that are used to go from the optimal solution to AP to the optimal solution to the ATSP. We have $\tilde C_1+\cdots+\tilde C_k\le\zeta^*$, and we have $\tilde C_e\ge 0$ for every edge $e$, so in particular, every $\tilde C_e\le\zeta^*$. 
\end{proof}
Now, though there are $n^2$ total edges, intuitively each vertex is on average likely incident to approximately $n\zeta^*=\zeta\log^2n$ edges that have reduced cost at most $\zeta^*$, giving a ``branching factor'' that is polylogarithmic in $n$ for the number of possible non-basic edges to choose. We will formalize this intuition in the following subsections.
\subsection{Specifying a possible ATSP}
A \textit{valid} cycle or path is a cycle or path respectively in $K_{A,B}$ whose edges outside of $T^*$ have length at most $2\z^*$. We know from Lemma \ref{AllTSPsmall} that the ATSP solution is formed from $M^*$ by taking  its symmetric difference  with valid cycles. For a valid cycle or path, let its \textit{non-basic cardinality} be the number of edges that are in the cycle or path but not in $T^*$.

By Lemma \ref{FewEdgesChanged}, all sets of cycles $Q_1,\ldots,Q_\ell$ that we need to consider along with $M^*$ to form the ATSP solution have non-basic cardinalities that total to at most $\log^{2+o(1)}(n)$.
\begin{lemma}\label{numvalid}
With high probability, for all $1\le k\le\log^3n$, there are at most $\log^{3k+1}n$ valid cycles with non-basic cardinality $k$.
\end{lemma}
\begin{proof}
Let $N_k$ be the number of valid cycles with non-basic cardinality $k$. Let $\tilde C_j,j=1,2,\ldots,k$ be the non-basic costs in a generic valid cycle. Then, if $m$ is as in Lemma \ref{basicpaths},
\[
\E(N_k)\le\binom mkk!2^k\Pr(\tilde C_j\le\zeta^*,1\le j\le k)\leq \binom mkk!2^k(\z^*)^k\leq (2\z\omega\log^2n)^k\leq \log^{3k}n.
\]
We choose $k$ alternating paths, order them in $k!2^k$ ways, and multiply by the probability that the reduced cost of the non-basic edge joining them into a cycle is at most $\z^*$.

The result then follows from the Markov inequality. 
\end{proof}

We also have an analogous statement for paths instead of cycles:
\begin{lemma}\label{numpaths}
With high probability, for all $1\le k\le\log^3n$, there are at most $n\log^{3k+2}n$ valid paths with non-basic cardinality $k$.
\end{lemma}
\begin{proof}
This is proven just like Lemma \ref{numvalid}, except that instead of choosing $k$ alternating paths, we can now choose $k+1$ alternating paths, giving an extra factor of $2\omega n\le n\log n$ throughout.
\end{proof}

Now, the previous two lemmas were not algorithmic. To prove Theorem \ref{th1}, we need to actually compute the valid cycles.
\begin{lemma}\label{findvalid}
With high probability, we can find and store all valid cycles and paths with non-basic cardinality at most $k$ in time $O(n^2\log^{3k+4}n)$.
\end{lemma}
\begin{proof}
We will prove this inductively. Our base case is that for $k=1$, we can take all of the $\omega n$ (see Lemma \ref{basicpaths}) basic alternating paths and see which pairs form a valid path or cycle with non-basic cardinality $1$. This takes $O(\om^2n^2\log n)$ time.

Assume that we have already found and stored all valid paths and cycles of non-basic cardinality less than $k$. Now, every valid path of non-basic cardinality $k$ is the concatenation of a valid path with non-basic cardinality $\lfloor\frac k2\rfloor$ and a valid path with non-basic cardinality $\lceil\frac k2\rceil$. So, we iterate over all pairs of valid paths, where the first has non-basic cardinality $\lfloor\frac k2\rfloor$ and the second has non-basic cardinality $\lceil\frac k2\rceil$. By Lemma \ref{numpaths}, the total number of such pairs is at most 
\[
\left(n\log^{3\lfloor k/2\rfloor+2}n\right)\left(n\log^{3\lceil k/2\rceil+2}n\right)\le n^2\log^{3k+4}n.
\]
We store the pairs where the end-point of the first equals the start point of the second, as these are exactly the valid cycles with non-basic cardinality $k$.

Finally, we iterate through the valid paths with non-basic cardinality $k$ to check which are cycles.
\end{proof}
\subsection{Iterating through Possible ATSP Solutions}\label{qaz}
First, we precompute and store all valid cycles with non-basic cardinality at most the value given by Lemma \ref{FewEdgesChanged}, which by Lemma \ref{findvalid} takes time at most 
\[
O\brac{\sum_{k=1}^{\log^{2+o(1)}n}n^2\log^{3k+4}n}=O(n^2\log ^{3\log^{2+o(1)}n+4}n)=O(e^{\log^{2+o(1)}n}).
\]
Then, we run through the possibilities for $k$, the number of edges in the ATSP but not in $T^*$. By Lemma \ref{FewEdgesChanged}, $k\le\log^{2+o(1)}n$ w.h.p.

Next, we run through the possibilities for $\ell$, the number of distinct valid cycles made by adding the $k$ edges into $T^*$, and accounting for the parity of the number of times a tree edge is in one of these cycles. We then specify the non-basic cardinalities $(k_1,\dotsc,k_\ell)$ of the $\ell$ valid cycles. Because $k_1+\dotsc+k_\ell=k$, we have that this step selects a partition of $k$, and thus we have (crudely) that there are at most $2^{2k}\le e^{\log^{2+o(1)}n}$ choices for $(k_1,\dotsc,k_\ell)$, and thus at most $e^{\log^{2+o(1)}n}$ possibilities to iterate through in the outer loop.

Now, for a fixed $(k_1,\dotsc,k_\ell)$, for each $1\le i\le\ell$, we specify the $i$th valid cycle by choosing one of the at most $\log^{3k_i+1}n$ pre-computed and stored valid cycles with non-basic cardinality $k_i$. Thus, the total amount of possibilities needed to iterate through for this particular $(k_1,\ldots,k_\ell)$ is at most 
\[
 \prod_{i=1}^\ell\log^{3k_i+1}n=\log^{3k+\ell}n\le e^{\log^{2+o(1)}n}
\]
as desired.

So we can iterate through each of these possible selections, check in polynomial time whether this gives a Hamilton cycle and if so evaluate its cost, and then remember the Hamilton cycle of minimum cost.

This finishes the proof of Theorem \ref{th1}, subject to the generalized proofs of Lemmas \ref{lem2a} and \ref{lemdiff} in Section \ref{properties}.
\subsection{Reducing the Space Complexity}
A downside of the previous algorithm is that of storing all $e^{\log^{2+o(1)}n}$ valid paths and cycles with non-basic cardinality up to $\log^{2+o(1)}n$ uses $e^{\log^{2+o(1)}n}$ memory. The previous algorithm can be amended to use only the optimal amount of space, $O(n^2)$ (the amount of space needed to store the costs) without significantly increasing the time complexity:
\begin{theorem}
With high probability, we can find the ATSP within $e^{\log^{2+o(1)}n}$ time and $O(n^2)$ space.
\end{theorem}
\begin{proof}
Instead of precomputing and storing all valid cycles with non-basic cardinality up to $\log^{2+o(1)}n$, we instead only precompute and store all valid paths and cycles with non-basic cardinality up to $\ell_0=\frac{\log n}{6\log\log n}$. By Lemma \ref{numvalid}, there are at most 
\[
O\left(\sum_{k=1}^{\ell_0}n\log^{3k+2}n\right)\le 2n\log^{\log n/(2\log\log n)+2}n\le 2n^{1.5}\log^2n
\] 
of these paths, each of which has length at most $\log^2n$ (as alternating paths in $T^*$ have length $O(\log n)$).

Now, as in the algorithm described in Section \ref{qaz}, we still iterate through all possible $(k_1,\dotsc,k_\ell)$. For the $i$th cycle, we specify it in one of two ways, depending on whether $k_i\le\ell_0$ or whether $k_i>\ell_0$.\\

\textbf{Case 1:} $k_i\le\ell_0$. Then just as before we iterate through the $O(\log^{3k_i+1}n)$ stored cycles.\\

\textbf{Case 2:} $k_i>\ell_0$. Now, we have not stored the valid cycles with non-basic cardinality $k_i$. Instead, we specify these cycles in a similar way to Lemma \ref{findvalid}. In particular, we can specify any cycle with non-basic cardinality $k_i$ by specifying $\lceil\frac{k_i}{\ell_0}\rceil$ valid paths of non-basic cardinalilty at most $\ell_0$ that concatenate to form this cycle. Using this process, as we had at most $2n^{1.5}\log^2n\le n^2$ valid paths of non-basic cardinality at most $\ell_0$, we have at most 
\[
n^{2(1+k_i/\ell_0)}=n^2e^{12k_i\log\log n}
\]
 possibilities to iterate through in order to iterate through every valid cycle of non-basic cardinality $k_i$.

Then the total amount of time this new lower-space algorithm takes on a given $(k_1,\dotsc,k_\ell)$ is asymptotically at most
\begin{align*}
&\prod_{i:k_i\leq \ell_0=1}\log^{3k_i+1}n \prod_{i:\ell_0<k_i\leq \log^{2+o(1)}n}n^2e^{12k_i\log\log n}\\
&\le(\log^{3k+\ell}n)(n^{\log^{2+o(1)}n/\ell_0}) e^{12k\log\log n}\\
&\le e^{\log^{2+o(1)}n}
\end{align*}
as desired.
\end{proof}
\section{Properties of the assignment problem}\label{properties}
In this section, we verify Lemmas \ref{lem2a} and \ref{lemdiff} under our more general distribution of costs. We assume that $C$ has a density $f$ and satisfies
\begin{enumerate}[(i)]
\item $f(x)=a+bx+O(x^2)$ for $0\leq x\leq L$, where $a,b$ are constants and $aL\geq 1$. 
\item (a) $\Pr(C\geq x)\leq \a e^{-\b x}$ for constants $\a,\b>0$, or (b) $f(x)=0$ for $x>L$.
\item To avoid some pathologies, we will also assume that there is a constant $M$ such that $f(x)\leq aM$ for $x\geq 0$.
\end{enumerate}
\subsection{$M^*$ only has low cost edges}
In this section we prove that w.h.p.,
\beq{2a}{
\text{$\max_{e\in M^*}\set{C(e)}\leq \g^*=\frac{\g\log n}{n}$ for some absolute constant $\g>0$.}
}
Much of this section is a direct adaptation of the proof used by Frieze and Sorkin to show the same lemma in the particular case where the cost distribution is $U[0,1]$ \cite{FS}.

Let $\dD_M$ etc. be as defined prior to \eqref{EP}. 
\begin{lemma}\label{cl1}
\Whp over random cost matrices $C$, for every perfect matching $M$, the (unweighted) diameter of $\dD_M$ is at most $k_0=\rdup{3\log_4n}$.
\end{lemma}
\begin{proof}
This is Lemma 5 of \cite{FS}.
\end{proof}

If we ignore the savings from edge deletions in traversing an alternating path then it follows fairly easily that 
\beq{lem2b}{
\text{$\max\set{C(i,j):\set{a_i,b_j}\in M}\leq \frac{\g_1\log^2n}{n}$ for some absolute constant $\g_1>0$.}
}
Indeed, for a fixed $i$ we have, where the density of $C$ is $a+bx+O(x^2)$ as $x\to0$,
\begin{align*}
\Pr\brac{C(i,j)\geq \frac{10\log n}{an}\text{ for }j\in[n/2]}&\leq \brac{1-\int_{x=0}^{10\log n/(an)}(a+bx+O(x^2))dx}^{n/2}\\
&\leq  \brac{1-\frac{10\log n}{n}+O\bfrac{\log^2n}{n^2}}^{n/2}=o(n^{-9}).
\end{align*}
It follows that w.h.p. all of the forward edges in the paths alluded to in Lemma \ref{cl1} have cost at most $\tfrac{10\log n}{an}$. If $x\in A$ and $y\in B$ then Lemma~\ref{cl1} implies that \whp\ there is a path from $x$ to $y$ for which the sum of the costs of the forward edges is at most $\tfrac{10k_0\log n}{an}$.  So if there is a matching edge of cost greater than $\tfrac{10k_0\log n}{an}$ then there is an alternating path using at most $k_0$ edges that can be used to give a matching of lower cost, contradiction. This verifies \eqref{lem2b}.

We now take account of the edges removed in an alternating path and thereby remove an extra $\log n$ factor. We will need the following inequality, analogous to Lemma 4.2(b) of~\cite{FG}, which deals with uniform $[0,1]$ random variables.
\begin{lemma}\label{lemFG}
Suppose that $k_1+k_2+\cdots+k_P=K\leq \k\log N,\k=O(1)$, and $Y_1,Y_2,\ldots,Y_P$ are independent random variables with $Y_i$ distributed as the $k_i$th minimum of $N$ independent copies of $C$. If $\m>1,\l=O(1)$ and $N$ is sufficiently large, then 
$$\Pr\left(Y_1+\cdots+Y_P\geq \frac{\m \k \log N}{N}\right)\leq N^{(a+\log \m-\th \l)\k},$$
where $\th=\b/2$. 
\end{lemma}
\begin{proof}
For $0\leq x\leq L$ we have $\Pr(C\leq x)=ax+O(x^2)\leq ax(1+Ax)$ for some $A>0$. Therefore, the density function $f_k(x)$ of the $k$th order statistic $Y_{(k)}$ satisfies
\begin{align*}
f_k(x)&=\binom{N}{k}(ax+O(x^2))^{k-1}a\brac{1-ax+O(x^2)}^{N-k}\qquad\text{for }x\leq L.\\
f_k(x)&\leq \binom{N}{k}(ax)^{k-1}aMe^{-\b(N-k)x}\qquad\qquad\qquad\qquad\ \ \qquad\text{for }x> L.
\end{align*}
Therefore the moment generating function of $Y_{(k)}$ satisfies
\begin{align*}
\E(e^{tY_{(k)}})&\leq a^kM\binom{N}{k}\int_{x\geq 0}e^{tx}(x+Ax^2))^{k-1}e^{-(N-k)\th x}dx\\
&\leq  a^kM\binom{N}{k}\int_{x\geq 0}x^{k-1}e^{-((N-k)\th-t-Ak)x}dx\\
&\leq \frac{a^kM N^k}{k((N-k)\th-t-A)^{k}}.
\end{align*}
So, if $Y=Y_1+\cdots+Y_P$ then 
\mults{
\E(e^{tY})\leq \prod_{i=1}^P\bfrac{a^{k_i} N^{k_i}}{((N-k_i)\th-t-A)^{k_i}}\leq\bfrac{aN}{\th N-t}^K\prod_{i=1}^P\brac{1+\frac{\th k_i+A}{(N-k_i)\th-t-k_i}}\\
\sim \bfrac{aN}{\th N-t}^K=\l^K,
}
if we take $t=(\th-a\l^{-1})N$.

So,
\[
\Pr\brac{Y\geq \frac{\m \k \log N}{N}}\leq \Pr\brac{e^{tY}\geq \exp\set{\frac{t\l\k\log N}{N}}}\le \frac{\l^K}{N^{(\th\l-a)\k}}.
\]
\end{proof}

We go along way towards proving \eqref{2a} by proving
\begin{lemma}\label{lem2ab}
The following holds with probability $1-o(n^{-2})$. For all $a\in A,b\in B$, $\vec D$ contains an alternating path from $a$ to $b$ of total cost less than $\g^*$
\end{lemma}
\begin{proof}
Let 
\begin{equation}\label{lexp}
Z_1=\max\left\{\sum_{i=0}^{k}C(x_i,y_i)-\sum_{i=0}^{k-1}C(y_i,{x}_{i+1})\right\},
\end{equation}
where the maximum is over sequences $x_0,y_0,x_1,\ldots,x_k,y_k$ where $(x_i,y_i)$ is one of the 40 shortest edges leaving $x_i$ for $i=0,1,\ldots,k\leq k_0=\rdup{3\log_4n}$, and $(y_i,x_{i+1})$ is a backwards matching edge. Also, in the maximum we assume that all $C(\cdot,\cdot)$ are bounded above  by $L=\tfrac{\g_1\log^2n}{n}$, see \eqref{lem2b}. We compute an upper bound on the probability that $Z_1$ is large. For any constant $\z>0$ we have
\begin{multline*}
\Pr\left(Z_1\geq \frac{\xi\log n}{n}\right)\lesssim\sum_{k=0}^{k_0} n^{2k+2}\frac{1}{(n-1)^{k+1}}\times\\
\int_{y=0}^L \left[\frac{1}{(k-1)!}\lfrac{y\log n}{n}^{k-1}
\sum_{\r_0+\r_1+\cdots+\r_{k}\leq 40(k+1)} q(\r_0,\r_1,\ldots,\r_{k};\xi+y)\right]dy
\end{multline*}
where 
$$
q(\r_0,\r_1,\ldots,\r_{k};\eta)  = \Pr\left(X_0+X_1+\cdots+X_{k}\geq \frac{\eta\log n}{n}\right),
$$
$X_0,X_1,\ldots,X_{k}$ are independent and $X_j$ is distributed as the $\r_j$th minimum of $n-1$ copies of $C$. (When $k=0$ there is no term $\frac{1}{k!}\lfrac{y\log n}{n}^{k}$).

{\bf Explanation:} 
We have $\leq n^{2k+2}$ choices for the sequence $x_0,y_0,x_1,\ldots,x_k,y_k$.
The term $\frac{1}{(k-1)!}\lfrac{y\log n}{n}^{k-1}dy$ asymptotically bounds the probability that the sum $\S=C(y_0,x_1)+\cdots+C(y_{k-1},x_{k})$ is in $\frac{\log n}{n}[y,y+dy]$. Indeed, if $C_1,C_2,\ldots,C_k$ are independent copies of $C$, then since $y\leq L$,
\begin{align*}
\Pr\brac{C_1+\cdots+C_k\in\frac{\log n}{n}[y,y+dy]}&=\int_{z_1+\cdots+z_k\in\frac{\log n}{n}[y,y+dy]}\prod_{i=1}^k\brac{1+O\bfrac{\log^2n}{n}}d\bz\\\
&\sim \int_{z_1+\cdots+z_k\in\frac{\log n}{n}[y,y+dy]}1d\bz=\frac{1}{(k-1)!}\lfrac{y\log n}{n}^{k-1}dy.
\end{align*}

We integrate over $y$. $\frac{1}{n-1}$ is the probability that $(x_i,y_i)$ is the $\r_i$th
shortest edge leaving $x_i$, and these events are independent for $0\leq i\leq k$. 
The final summation bounds the probability that the associated edge lengths sum to at least $\frac{(\xi+y)\log n}{n}$.

It follows from Lemma \ref{lemFG} that if $\g$ is sufficiently large then, for all $y\geq 0$,
$q(\r_1,\ldots,\r_k;\g+y)  \leq n^{-(\g+y)/2}$ and since the number of choices for $\r_0,\r_1,\ldots,\r_k$ is at most $\binom{41k+40}{k}$ (the number of non-negative integral solutions to $x_0+x_1+\ldots+x_{k+1}=40(k+1)$) we have
\begin{align*}
\Pr\left(Z_1\geq \g^*\right)& \leq
2n^{2-\g/2}\sum_{k=0}^{k_0}\frac{\log^{k-1}n}{(k-1)!}\binom{84k}{k}\int_{y=0}^\infty y^{k-1}n^{-y/2}dy\\ 
&\leq 
2n^{2-\g/2}\sum_{k=0}^{k_0}\frac{\log^{k-1}n}{(k-1)!}\lfrac{168e}{\log n}^{k}\G(k)\\ 
&\leq 2n^{2-\g/2}(168e)^{k_0+1}\\
&=o(n^{-2}).
\end{align*}
\end{proof}

{\bf Proof of Lemma \ref{lem2a}}\\
Suppose $e=\set{a_i,b_j}\in M^*$ and $C(e)>\g^*$. It follows from Lemma \ref{lem2ab}  that there is an alternating path $P=(a_1,\ldots,b_j)$ of cost at most $\g^*$. But then deleting $e$ and the $M^*$-edges of $P$ and adding the non-$M^*$ edges of $P$ to $M^*$ creates a matching from $A$ to $B$ of lower cost than $M^*$, contradiction.

\section{From exponential mean one to more general distributions}\label{generalizedist}
\subsection{A high probability bound on $Z_{\ATSP}-Z_{\AP}$}
We now verify \eqref{diff} with our more general distribution for costs.  We let the  ${\widehat C}(i,j)$ be independent copies of a uniform $[0,1]$ random variable and then let $C(i,j)=F^{-1}({\wh C}(i,j))$, where $F(x)=\Pr(C\leq x)$. \\
(Note that $\Pr(C(i,j)\leq x)=\Pr(F^{-1}({\wh C}(i,j))\leq x)=\Pr(\wh C(i,j)\leq F(x))=F(x)$ and $C(i,j)=(1+O(x))\wh C(i,j)$ for $x=o(1)$.)

Then, using  Lemma \ref{lem2a}, we have
\begin{align*}
C(ATSP)&\leq \brac{1+O\bfrac{\log n}{n}}{\wh C}(ATSP)\\
&\leq \brac{1+O\bfrac{\log n}{n}}\brac{{\wh{C}(AP)+O\bfrac{\log^2n}{n}}},\qquad\text{from \eqref{diff}},\\
&\leq {\wh C}(AP)+O\bfrac{\log^2n}{n}\\
&\leq C(AP)+O\bfrac{\log^2n}{n}.
\end{align*}
\subsection{Dealing with Lemma \ref{steady}}
Lemma \ref{steady} is only valid for the costs $C$ being exponential with mean 1. However, we will now show that our result still holds on a more general class of distributions. If we replace the costs $C(i,j)$ by $\widehat C(i,j)=aC(i,j)$ then our proof will go through almost unchanged. This yields a distribution as in the introduction where $a$ is now 1. We assume then that $a=1$.

Suppose that the edge costs $C=C(i,j)$ are distributed as claimed in Theorem \ref{th1}. We can naturally couple these edge costs to EXP(1) edge costs, which we will call $X$, as follows: first sample a cost $c$ from $C$, then find $p=\Pr(C\le c)=F(c)$, and set the cost $x$ under $X$ such that $1-e^{-x}=p$ or $x=\log\tfrac{1}{1-F(c)}$ (in other words, drawing from the same places on the respective cumulative distribution functions).

We know $\Pr(C\leq z)=z+O(z^2)$ and $\Pr(X\leq z)=z+O(z^2)$ as $z\to 0$. We also know from Lemma \ref{AllTSPsmall} that under $X$, w.h.p.~all edges in the AP solution have cost at most $2\z^*=O\bfrac{\log^2 n}{n}$. Suppose now that the costs of the edges in the optimal ATSP under $X$ are $U_i$, $i=1,2,\ldots,n$ where each $U_i\leq 2\z^*$. Now we know that $\sum_{i=1}^nU_i=O(1)$ w.h.p. \cite{FS}, which then implies that $\sum_{i=1}^nU_i^2=O\bfrac{\log^2n}{n}$. (We maximise the sum by putting $U_i=2\z^*$ for at most $O(n/\log^2n)$ indices and putting $U_i=0$ for the remaining indices.) It follows that $C(ATSP)-X(ATSP)\le O\bfrac{\log^2n}{n}$. The same argument (noting Lemma \ref{lem2a} holds for both $X$ and $C$) gives $|X(AP)-C(AP)|\le O\bfrac{\log n}{n}$. Therefore, $|C(ATSP)-X(AP)|\le O\bfrac{\log^2n}{n}$, so our enumeration above (starting from the optimal AP under $X$ and replacing $\z^*$ by $\tfrac{K\log^2n}{n}$ for sufficiently large $K$) will also w.h.p.~find the optimal ATSP under $C$.

\section{Summary and open questions}\label{sum}
One can easily put the enumerative algorithm in the framework of branch and bound. At each node of the B\&B tree one branches by excluding edges of $M^*$. So, at the top of the tree the branching factor is $n$ and in general, at level $k$, it is $n-k$. W.h.p. the tree will have depth at most $e^{\log^{2+o(1)}n}$.

The result of Theorem \ref{th1} does not resolve the question as to whether or not there is a branch and bound algorithm that solves ATSP w.h.p. in polynomial time. This remains an open question.

Less is known probabilistically about the symmetric TSP. Frieze \cite{F} proved that if the costs $C(i,j)=C(j,i)$ are independent uniform $[0,1]$, then the asymptotic cost of the TSP and the cost 2F of the related 2-factor relaxation are asymptotically the same. The probabilistic bounds on $|TSP-2F|$ are inferior to those given in \cite{FS}. Still, it is conceivable that the 2-factor relaxation or the subtour elimination constraints are sufficient for branch and bound to run in polynomial time w.h.p.

Yatharth Dubey \cite{Dub} pointed out that combining the above analysis with arguments from \cite{DDM} shows that using the subtour elimination LP relaxation in a branch and bound algorithm will also lead to a quasi-polynomial time algorithm w.h.p.

\end{document}